\documentclass[reprint, superscriptaddress, showkeys, prmaterials]{revtex4-1}
\usepackage{amsmath,amssymb}
\usepackage[english]{babel}
\usepackage[utf8]{inputenc}
\usepackage[colorinlistoftodos, color=green!40, prependcaption]{todonotes}
\usepackage{amsthm}
\usepackage{mathtools}
\usepackage{xcolor}
\usepackage{graphicx}
\usepackage[left=23mm,right=13mm,top=35mm,columnsep=15pt]{geometry} 
\usepackage{adjustbox}
\usepackage{placeins}
\usepackage[T1]{fontenc}
\usepackage{lipsum}
\usepackage{csquotes}
\usepackage[amssymb,thinqspace,thinspace]{SIunits}
\usepackage{color}
\usepackage{tabularx}
\usepackage[modulo]{lineno}
\usepackage{natbib}
\usepackage{nicefrac}
\usepackage{mathtools}\usepackage[pdftex, pdftitle={Article}, pdfauthor={Author}]{hyperref} 
\begin{document}
\title{Interface characteristics in an $\alpha$+$\beta$ titanium alloy}

\author{Abigail K. Ackerman}
    \email{Corresponding author: a.ackerman14@imperial.ac.uk}
    \affiliation{Department of Materials, Royal School of Mines, Imperial College London, Prince Consort Road, London, SW7 2BP, UK}
\author{Vassili A. Vorontsov}
    \affiliation{Department of Materials, Royal School of Mines, Imperial College London, Prince Consort Road, London, SW7 2BP, UK}
\author{Ioannis Bantounas}
    \affiliation{Department of Materials, Royal School of Mines, Imperial College London, Prince Consort Road, London, SW7 2BP, UK}
\author{Yufeng Zheng}
    \affiliation{Center for the Accelerated Maturation of Materials and Department of Materials Science and Engineering, The Ohio State University, Columbus, OH 43212, USA}
    \affiliation{Department of Chemical and Materials Engineering, University of Nevada Reno, Reno, NV 89557, USA}
\author{Yanhong Chang}
    \affiliation{Max-Planck-Institut f\"{u}r Eisenforschung GmbH, Max-Planck-Str. 1, 40237 D\"{u}sseldorf, Germany}
\author{Thomas McAuliffe}
    \affiliation{Department of Materials, Royal School of Mines, Imperial College London, Prince Consort Road, London, SW7 2BP, UK}
\author{William A. Clark}
    \affiliation{Center for the Accelerated Maturation of Materials and Department of Materials Science and Engineering, The Ohio State University, Columbus, OH 43212, USA}
\author{Hamish L. Fraser}
    \affiliation{Center for the Accelerated Maturation of Materials and Department of Materials Science and Engineering, The Ohio State University, Columbus, OH 43212, USA}
\author{Baptiste Gault}
    \affiliation{Max-Planck-Institut f\"{u}r Eisenforschung GmbH, Max-Planck-Str. 1, 40237 D\"{u}sseldorf, Germany}
    \affiliation{Department of Materials, Royal School of Mines, Imperial College London, Prince Consort Road, London, SW7 2BP, UK}
\author{David Rugg}
    \affiliation{Rolls-Royce plc., Elton Road, Derby, DE24 8BJ, UK}
\author{David Dye}
     \email{david.dye@imperial.ac.uk}
    \affiliation{Department of Materials, Royal School of Mines, Imperial College London, Prince Consort Road, London, SW7 2BP, UK}
   
\date{\today}

\begin{abstract}
The $\alpha$/$\beta$ interface in Ti-6Al-2Sn-4Zr-6Mo (Ti-6246) was investigated \emph{via} centre of symmetry analysis, both as-grown and after 10\% cold work. Semi-coherent interface steps are observed at a spacing of 4.5 $\pm$1.13 atoms in the as-grown condition, in good agreement with theory. Lattice accommodation is observed, with elongation along [$\bar{1}$2$\bar{1}$0]$_\alpha$ and contraction along [10$\bar{1}$0]$_\alpha$. Deformed $\alpha$ exhibited larger, less coherent steps with slip bands lying in $\{110\}_\beta$. This indicates dislocation pile-up at the grain boundary, a precursor to globularisation during heat treatment. Atom probe tomography measurements of secondary $\alpha$ plates in the fully heat treated condition showed a Zr excess at the interface, which was localised into regular structures indicative of Zr association with interface defects, such as dislocations. Such chemo-mechanical stabilisation of the interface defects would both inhibit plate growth during elevated temperature service and the interaction of interface defects with gliding dislocations during deformation.    
\end{abstract}


\maketitle


\section{Introduction}
\label{Intro}
Titanium alloys are widely used in safety-critical aerospace applications due to their exceptional specific fatigue-allowable strengths~\cite{Titanium}. The solid-state transformation from the high temperature \emph{bcc} $\beta$ phase to the \emph{hcp} $\alpha$ phase offers the opportunity for microstructural tailoring through solid-state processing, which allows fine-grained microstructures to be produced~\cite{Lutjering1999}. The orientation relationship between the phases obtained by minimising the interface strain and maximising interface coherency is approximately \textcolor{red}{$\{110\}_\beta \! \parallel \! \{0002\}_\alpha$ $\langle 111\rangle_\beta\!\parallel\!\langle 11\bar{2}0\rangle_\alpha$~\cite{Furuhara1996}.} In $\alpha-\beta$ alloys, naturally-grown $\alpha$ forms as plates nucleated from pre-existing $\alpha$ or from $\beta$ grain boundaries, \emph{e.g.} in a Widmanst\"{a}tten morphology; equiaxed $\alpha$ can then be produced by hot working and globularisation~\cite{Stefansson2002, Park2012}.

The low energy, broad face habit plane of $\alpha$ plates is commonly held to be $\{112\}_\beta \parallel \{10\bar{1}0\}_\alpha$~\cite{Titanium}, although other works have also reported $\{11\, 11\, \overline{13}\}_{\beta}\! \parallel\! \{\overline{27}\, 20\, 7\, 0\}_\alpha$~\cite{Furuhara1991}. Detailed Transmission Kikuchi Diffraction (TKD) analysis has recently observed both possibilities in the same specimen~\cite{Tong2017}. In order to accommodate the interface misfit between the two phases, the plates contain interface defects, including interfacial dislocations~\cite{Dc1989,Furuhara1991a}, structural ledges~\cite{Shiflet1998,Cabibbo2013} and misfit-compensating ledges. These give rise to plate thickening through the classical terrace-ledge-kink model, where the ledges are incoherent and therefore mobile whereas the terraces are coherent and therefore immobile. Pond and Hirth~\cite{Pond2003,Hirth2007} analysed the ledges as being composed of both dislocation character and a step associated with the core, terming the overall defect a \emph{disconnection}. 

Models of the $\alpha$/$\beta$ interface have therefore been constructed to predict the interface, based on both the O-lattice theory of Bollmann~\cite{Bollmann} and the topological model of Hirth and Pond~\cite{Hirth1996,Pond2003}, with the topological approach generally finding more applications. Recently, Zheng \emph{et al.}~\cite{Zheng2018} examined the interfaces formed in refined Ti-5553, validating the calculated disconnection spacing and direction in the interfaces observed.

The disconnection character and spacing is of technological significance because these will determine the mobility of the interface and therefore the plate thickening rates that arise. This will in turn be important to the plate formation kinetics. Therefore these defects determine refinement of the microstructures achieved by heat treatment, given the mobilities of the rate-controlling solutes at the $\alpha$/$\beta$ interface. It can also be hypothesised that the interface defects and their mobility are in some way related to the nucleation of branching side-plates where fine-scale secondary $\alpha$ are formed in heavily $\beta$-stabilised alloys such as Ti-6Al-2Sn-4Zr-6Mo (Ti-6246).  

Another aspect that is likely critical to interface mobility is local chemical composition and the possible segregation of particular species to such an interphase or grain boundary interface. The general subject of segregation to interfaces was discussed, for instance, by Raabe \emph{et al.}~\cite{Raabe2014}, where the authors describe how the decoration of grain boundaries by solutes or secondary phases can be used to benefit the performance of materials. Atom probe tomography (APT) is often employed in such investigations~\cite{Breen2016,Herbig2014}. In recent years, there have been many reports of such segregation to interfaces in materials other than titanium, such as aluminium. For example Pandey \emph{et al.}~\cite{Pandey2019} observed Zr segregation to the eutectic Al$_{3}$Ni/$\alpha$-Al interface in a Al rich Al-Ni eutectic alloy with the addition of 0.15 wt.\% Zr. It was suggested that the presence of Zr at this interface reduces the interfacial energy. Solute segregation has also been observed in a model Al-Zn-Mg-Cu alloy~\cite{Zhao2018}, where 10$\usk\nano\meter$ precipitate-free zones had formed adjacent to enriched grain boundaries. Zn is frequently depleted at Al grain boundaries, in regions 7-15$\usk\nano\meter$ wide~\cite{Sha2018}. It has also been proposed that oxygen may pin the grain boundary in nanocrystalline Al~\cite{Tang2012}. In steels, Miyamoto \emph{et al.}~\cite{MIYAMOTO2018168} observed boron segregation in a low carbon steel, the extent of which reduced at low angle grain boundaries. In an Fe(Cr) nanocrystalline alloy, solute segregation was found to be dependent on grain boundary type~\cite{Zhou2016}. However, grain and phase boundary solute segregation has not yet received a similar level of attention in Ti alloys.

In this article, we investigate the interface of primary $\alpha$ plates formed in Ti-6246 with basketweave primary $\alpha$, in order to examine the generality of the findings of Zheng \emph{et al.}~\cite{Zheng2018}. We then examine the effect of deformation on the interface structure \textcolor{red}{, and discuss the implications of these changes for the subsequent evolution of the microstructure during processing to produce fine grained primary $\alpha$ in commercial product}. In addition, atom probe tomography data are presented that measure chemical segregation in the vicinity of $\alpha$ interfaces in the aged condition. We finally discuss how our findings on the deformation and segregation might then facilitate \emph{e.g.} the nucleation of side-plates and the remodelling of $\alpha$ plates during globularisation.

\section{Experimental methods}
\label{Experimental}
Specimens were prepared from a high pressure compressor disc of nominal composition Ti-6Al-2Sn-4Zr-6Mo (wt.\%), \textcolor{red}{Ti-10.8Al-0.8Sn-2.1Zr-3.0Mo (at.\%)} supplied by Rolls-Royce plc. Derby.  A $10\times 10 \times 10\usk\milli\meter$ sample was heat treated for $30\usk\minute$ at $960\celsius$ then cooled at $7\celsius\usk\minute$$^{-1}$ to $800\celsius$, followed by manual water quenching. The resulting microstructure featured primary $\alpha$ laths within a $\beta$ matrix, to allow a clean interface to be investigated via electron microscopy. The material was then prepared using a standard metallographic process. Specimens were then etched using Kroll's solution ($100\usk\milli\litre$ H$_2$O, 6$\usk\milli\litre$ HNO$_3$, 3$\usk\milli\litre$ HF), to image grain boundaries in the scanning electron microscope (SEM), prior to lift-out. 

To produce a deformed microstructure, a bar $15\times 15 \times 120\usk\milli\meter$ was flat rolled at room temperature to 90$\%$ of its starting thickness in one pass. A section cut from the centre of the bar was then prepared for electron backscatter diffraction (EBSD). EBSD orientation mapping was performed using a Zeiss Auriga FEG-SEM with an Bruker HKL eFlash EBSD detector to locate grains for ion milling, using an FEI Helios NanoLab 600 DualBeam system with an Omniprobe\textsuperscript{TM} micromanipulator using a standard lift-out procedure~\cite{Gi1999}. Scanning transmission electron microscopy (STEM) was performed using a probe-corrected FEI Titan$^3$\textsuperscript{TM} 80-300.

Lattice parameters were found by X-ray diffraction (XRD) using a Panalytical X’Pert Pro diffractometer system, with Cu K$_\alpha$ radiation. Spectrum patterns were recorded in the range 5-80$\degree$ 2$\theta$, with a step size of 0.0334$\degree$. Peak fitting was performed using Topas-Academic.

To characterise the interface structure, image masks were placed over the FFT pattern spots to refine the image, using Gatan Digital Micrograph software. Atom column positions were calculated using a MATLAB script to find nearest neighbour coordinates for symmetry analysis

As-received, fully heat treated needle-shaped specimens for APT were prepared by the cryogenic focused ion beam (cryo-FIB) milling protocol described by Chang et al.~\cite{Chang2019} to prevent the undesired H pick-up from the environment. Site-specific lift-outs were conducted with a 30 kV, 6–9 nA Xenon plasma source at ambient temperature, and the subsequent annular milling was done using a 30 kV, 0.46 nA to 24 pA Xe beam after the stage was cooled to $\sim$135$\degreecelsius$ using the setup described in~\cite{Stephenson2018}. The final milling was conducted with a 2 kV, 24 pA Xe ion beam under cryogenic conditions. APT measurements were performed using a Cameca LEAP 5000 XR operated in high-voltage pulsing mode with 20\% pulse fraction, 250 kHz pulse frequency and a target detection rate of 5 ions per 1000 pulses at a base temperature of 50 K. The pressure in the ultra-high vacuum chamber was consistently below $4 \times 10^{-9}$ Pa. \textcolor{red}{Isocomposition surfaces~\cite{Hellman2003} of 2 at.\% were then used to analyse Zr concentration at the interface, where a set of 3D surfaces delineate regions within the point cloud that contain over 2 at.\% Zr. }

\section{Modelling}
\subsection{Phenomenological Theory of Martensitic Transformations (PTMT)}

\begin{figure*}[t!]
\centering\includegraphics[width=1\linewidth]{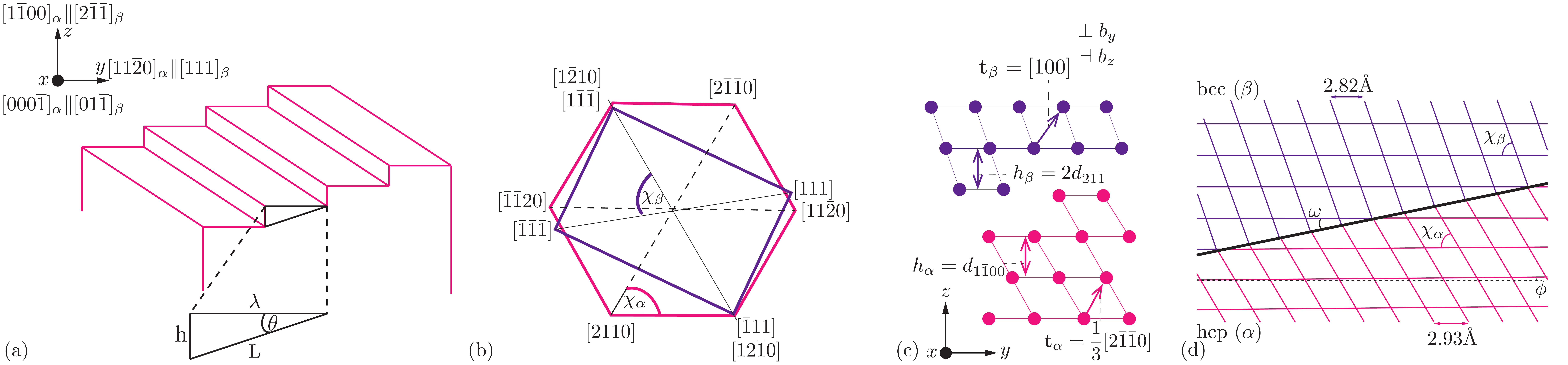}
\caption{(a) Schematic illustration of the broad interface between the $\alpha$ and $\beta$ phases, with the corresponding (ideal) axes in each phase. (b) The cubic lattice (purple) and the hexagonal lattice (pink), illustrating the Burgers orientation relationship. There is a fixed angle $\chi$ of each lattice, and lattice parameter, a. (c) The trigonometric relation of the interface parameters, and (d) the lattice matching defining the angles $\omega$ and $\phi$.}\label{fig:modelref}
\end{figure*}

According to the phenomenological theory of martensite crystallography/transformations (PTMC/PTMT)~\cite{Pond2003}, a deformation, $\mathbf{S}$, transforms the cubic $\beta$ lattice into the hexagonal $\alpha$ lattice at the Burgers orientation;
\begin{equation}
\label{eq:S}
\begin{split}
\mathbf{S}  = \Omega  & 
\begin{bmatrix}
\,\,\,\,\,\Omega^{-1} & 0 & 0\\
0 & 1 &  (\cos\chi_\beta - \cos\chi_\alpha)/\sin\chi_\beta\,\, \\
0 & 0 & \sin\chi_\alpha/\sin\chi_\beta 
\end{bmatrix}\\ = &
\begin{bmatrix}
1& 0& 0\\
0& 1.0420 & -0.1842\\
0&0&0.9572\\
\end{bmatrix}
\end{split}
\end{equation}
where $\Omega = 2a_\alpha/\sqrt{3}a_\beta$ and $a_\alpha$ and $a_\beta$ are the lattice parameters of each phase, obtained by X-ray diffraction. These were measured to be \textcolor{red}{$a_{\alpha}$ = 0.29346$\pm$0.0004$\usk\nano\meter$, $c_{\alpha}$ = 0.4693$\pm$0.0006$\usk\nano\meter$ and $a_{\beta}$ = 0.32519$\pm$0.0004$\usk\nano\meter$}. No deformation is assumed along the $x$ axis (\emph{hcp} $c$ direction) \textcolor{red}{as the coherency strain is small in this direction, so the lattice invariant deformation (LID) does not need to be considered~\cite{Bywater1972, Pond2003}}, and therefore this can be treated as a 2-dimensional problem. $\chi_\beta = 70.53\degree$ and $\chi_\alpha = 60\degree$ are the angles between lattice planes, which are set for \emph{hcp} and \emph{bcc}, Figure 1(b and c). Wayman~\cite{Wayman} decomposed the deformation tensor $\mathbf{S}$ into $\mathbf{R}$$_1$$\mathbf{B}$, where $\mathbf{B}$ is the lattice transformation (Bain strain, \textcolor{red}{which must be symmetric}) and $\mathbf{R_1}$ is a rigid body rotation. 
\begin{equation}
\label{eq:B}
\mathbf{B}=
\begin{bmatrix}
1& 0& 0\\
0& 1.0376 & -0.0956\\
0&-0.0956&0.9700\\
\end{bmatrix}
\end{equation}

$\mathbf{R}_1$ is then:
\begin{equation}
\begin{split}
\label{eq:R1}
\mathbf{R}_1 & = \mathbf{S}\mathbf{B}^{-1} = \mathbf{S}\left(\mathbf{S}^{\text{T}}\mathbf{S}\right)^{-\frac{1}{2}}\\
& =
\begin{bmatrix}
1& 0& 0\\
0& 0.9958 & -0.0918\\
0&0.0918&0.9958\\
\end{bmatrix}
\end{split}
\end{equation}
%
The matrix $\mathbf{B}$ is multiplied by an invariant vector in the invariant plane, $\mathbf{v}$. $\mathbf{v\cdot Bv}$ can be used to find the angle of rotation, $\textcolor{red}{\psi_P}$, where $\mathbf{Bv}$ is the undistorted but rotated vector. To discover the orientation of the invariant plane, solutions are found for equation~\ref{eq:v}, \textcolor{red}{ where $\mathbf{v'}$ is the transpose of the vector \textbf{v}.}

\begin{equation}
\label{eq:v}
    \mathbf{v'B^{2}v} = \mathbf{v'v}
\end{equation}

As the unit vector $\mathbf{v}$ is in the $y-z$ plane, $x=0$, equation~\ref{eq:v} becomes:

\begin{equation}
\begin{split}
    y^2(B_{22}^2+B_{23}B_{32})+2yz(B_{22}B_{23}+B_{23}B_{33})+
    \\z^2(B_{23}B_{32}+B_{33}^2)=y^2+z^2
\end{split}
\end{equation}

\textcolor{red}{where $B_{23} = B_{32}$}. Values from equation~\ref{eq:B} can now be inserted. Additionally, it can then be said that $y^2 + z^2$ = 1 as $\mathbf{v}$ is a unit vector so that:

\begin{equation}
    0.0793(1-z^2) - 0.0556z^2 - 0.3816z(\sqrt{1-z^2}) = 0
\end{equation}

This can then be solved numerically to give:

\begin{equation}
    \mathbf{v} = \left(\begin{array}{c} 0 \\ 0.9771 \\ 0.2141 \end{array}\right)
\end{equation}

The angle of inclination of the invariant plane, IP, with respect to the terrace is then given by:

\begin{equation}
    \textcolor{red}{\omega_P} = \tan^{-1}(v_{y}/v_{z})=12.27\usk\degree
\end{equation}
where $v_{y}$ and $v_{z}$ are the $y$ and $z$ components of $\mathbf{v}$.

The rotation between $\mathbf{v}$ (invariant vector) and $\mathbf{Bv}$ (rotated but undistorted vector) is denoted $\mathbf{R_{2}}$. To calculate this and the accompanying angle of rotation, a standard rotation matrix is used \textcolor{red}{for the angle between $\mathbf{v}$ and $\mathbf{Bv}$, $\psi_P$}. 
\begin{equation}
\label{eq:theta}
\textcolor{red}{\psi_P} = \cos^{-1}\left(\frac{\mathbf{v}\cdot \mathbf{Bv}}{\|\mathbf{v}\| ~ \|\mathbf{Bv}\|}\right) 
= 5.7981^{\circ}
\end{equation}
This angle can be used to calculate the rotation $\mathbf{R_2}$, making $\mathbf{v}$ the invariant vector, where $\mathbf{R_2Bv}=\mathbf{v}$.
\begin{equation}
\begin{split}
\mathbf{R_2} & = 
\begin{bmatrix}
1 & 0 & 0\\
0 & \cos(\psi) & -\sin(\psi) \\
0 & \sin(\psi) & \cos(\psi)
\end{bmatrix}\\
& =
\begin{bmatrix}
1& 0& 0\\
0& 0.9949 & -0.1010\\
0&0.1010&0.9949\\
\end{bmatrix}
\end{split}
\end{equation}
%
$\mathbf{R}_3 = \mathbf{R}_2\mathbf{R}^{-1}_{1}$ is the overall misorientation of the $\alpha$ lattice away from the Burgers orientation, with a rotation angle $\textcolor{red}{\phi_P} = \cos^{-1}(\text{R}_{3yy})$, Figure 1(d). 

The final rotation, $\mathbf{R_3}$, can now be calculated, where:

\begin{equation}
\mathbf{R_3} = \setlength\arraycolsep{2pt}
\mathbf{R_{2}R_{1}^{-1}}
=
\begin{bmatrix}
1& 0& 0\\
0& 0.99996 & -0.0.0093\\
0&0.0093&0.99996\\
\end{bmatrix}
\end{equation}

This gives the final angle of rotation, the rigid body rotation of the $\alpha$ lattice away from the Burgers orientation, as 0.5331\degree.


\subsection{Topological theory}
Alternatively, the relationship between the phases can be calculated using a topological model~\cite{Pond2003}, where the $\beta$ lattice is stretched and the $\alpha$ lattice is compressed by $\varepsilon_{yy}/2$ along $\langle 111\rangle_\beta\!\parallel\!\langle 11\bar{2}0\rangle_\alpha$.

Now, the step heights of the $\alpha$ and $\beta$ phases can be calculated using lattice parameters and angles outlined,
\begin{equation}
\begin{split}
\label{eq:h}
h_\alpha &= a_\alpha \sin\left(\chi_\alpha\right)=0.2541\;\mathrm{nm}\\
h_\beta &= \frac{\sqrt{3}a_\beta}{2}\sin\left(\chi_\beta\right)=0.2655\;\mathrm{nm}
\end{split}
\end{equation}
The Burgers vector of the disconnections is given by
$\mathbf{b} = \mathbf{t}_\beta - \mathbf{t}_\alpha$ with components $b_z = h_\beta-h_\alpha$ (see Figure 1(c) for definitions) and
\begin{equation}
b_y = 3^{-0.5}a_\beta(1+\varepsilon_{yy}/2) - 0.5a_\alpha(1-\varepsilon_{yy}/2) 
\end{equation}
Pond \emph{et al}.~\cite{Pond2003} show that 
\begin{equation}
\varepsilon_{yy} = \frac{b_y \tan\theta+b_{z}\tan^2\theta}{h_\alpha} = 0.0412\;\mathrm{nm}
\end{equation}
For this system, $b$ = 0.0492 nm, $b_y$ = 0.0479 nm and $b_z$ = 0.0114 nm.
This quadratic equation can be solved for $\tan\theta$ (Figure 1(a)). Then, $\textcolor{red}{\omega_T = \theta_T - \phi_T}$, Figure 1(d). $\textcolor{red}{\phi_T}$ is then calculated as follows:
\begin{equation}
\begin{split}
\textcolor{red}{\phi_T} &= 2\sin^{-1}\left(\frac{(b_z\cos\theta-b_y\sin\theta)\sin\theta}{2h_\alpha} + \frac{\varepsilon_{yy}\sin\theta\cos\theta}{2}\right)\\
&= 0.5108\degree
\end{split}
\end{equation}
Thus, one obtains the disconnection spacing, $\textcolor{red}{L_T}$ and the disconnection spacing along the terrace plane, $\textcolor{red}{\lambda_T}$:
\begin{align}
\textcolor{red}{L_T} &= \frac{h_\alpha}{\sin\theta} = 1.2473\;\mathrm{nm} \\
\textcolor{red}{\lambda_T} &= \textcolor{red}{L_T}\cos\theta = 1.2211\;\mathrm{nm}
\end{align}
\section{Results}
\label{results}
\subsection{Undeformed $\alpha$/$\beta$ interface}
\begin{figure}[t]
\centering\includegraphics[width=\linewidth]{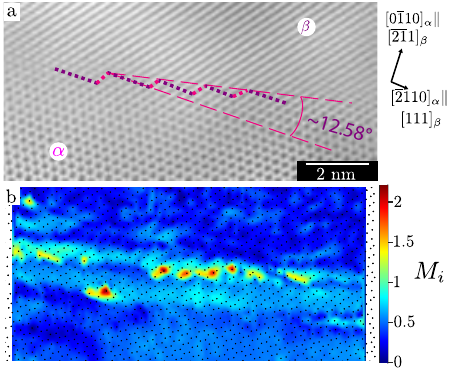}
\caption{(a) HAADF HR-STEM of the $\alpha$/$\beta$ interface. The top of the image is viewing the (110) $\beta$ plane, the bottom is the (0002) $\alpha$ plane. The stepped interface can be seen in the middle, (b) colour map of $M_i$, indicating where the positions of steps are along the interface through deviation from symmetry.}\label{fig:NGalpha}
\end{figure}

The $\alpha$/$\beta$ interface was imaged using HR-STEM, Figure~\ref{fig:NGalpha}. The top of the image is viewed along [110]$_\beta$, and the bottom along [0002]$_\alpha$. Foil heterogeneity caused deviation from the zone axis in some parts of the image, reducing the definition of some atomic columns. To address this, the image FFT was filtered via pattern spot masking, Lucy-Richardson deblurring and Gaussian point spread function (PSF) re-convolution. The nearest neighbours could then be calculated for each column enabling analysis of the local crystal symmetry to estimate interfacial step locations, Figures~\ref{fig:NGalpha}(b) and ~\ref{fig:DFalpha}(d). The interfacial symmetry parameter $M_i$ mapped in Figure~\ref{fig:NGalpha}(b) is given by:
\begin{equation}
M_i=\left(||\textstyle{\sum_{n=1}^{3}}\mathbf{r}_n||\!\cdot\!||\textstyle{\sum_{n=1}^{2}}\mathbf{r}_n||\right)/\,||\mathbf{r}_1||^2
\end{equation}
where $\mathbf{r}_n$ is the relative position vector of the nearest neighbour of rank $n$ for a given atomic column taken as the origin, \emph{i.e.} $\mathbf{r}_1$ is the first nearest neighbour position. The crystal symmetry of the $\alpha$ and $\beta$ phases gives non-zero values of $M_i$ only at the interface, due to \emph{e.g.} crystallographic defects and regions with non-uniform lattice strain. Hence, \textcolor{red}{$M_i$} gives a qualitative measure of deviation from the ideal structure in the vicinity of the interface.

Multiple disconnection spacings were measured from the image giving an average disconnection spacing $\lambda$ of 4.5 $\pm$1.13 atoms (1.5nm) and an overall inclination $\omega$ of $12.6\pm1^{\circ}$, allowing comparison with the PTMC and topological models. 
The topological model predicted a disconnection spacing $\lambda_T$ of 4.25 atoms (1.22$\usk$nm), with $\omega_T$=12.29$^{\circ}$ and $\phi_T$=0.5337$^{\circ}$, \textcolor{red}{whilst the PTMC model predicted} $\omega_P$=$12.27^\circ$ and $\phi_P$=0.5331$^{\circ}$. \textcolor{red}{Good agreement between both models and the measurements are obtained, similar to the result obtained by Zheng et al for alloy Ti-5553~\cite{Zheng2018}, who only compared to the topological model.}


In addition to disconnection spacing $\lambda$, symmetry analysis highlighted a distorted region in the $\alpha$ phase adjacent to the interface (light blue). Its depth can be observed to be $\approx$1nm and within it the $\alpha$ lattice shows a gradual dimensional distortion in the course of deforming into the $\beta$ lattice. It exhibits elongation and contraction along the $[\bar{1}2\bar{1}0]_{\alpha}$ and $[10\bar{1}0]_{\alpha}$ crystallographic directions, respectively. Given the very thin specimen geometry this effect is unlikely to result from inclination of the interface relative to the beam direction and can therefore be assumed to be the accommodation of the new lattice form.

\subsection{Interface chemistry}
\begin{figure*}[t!]
\centering\includegraphics[width=\linewidth]{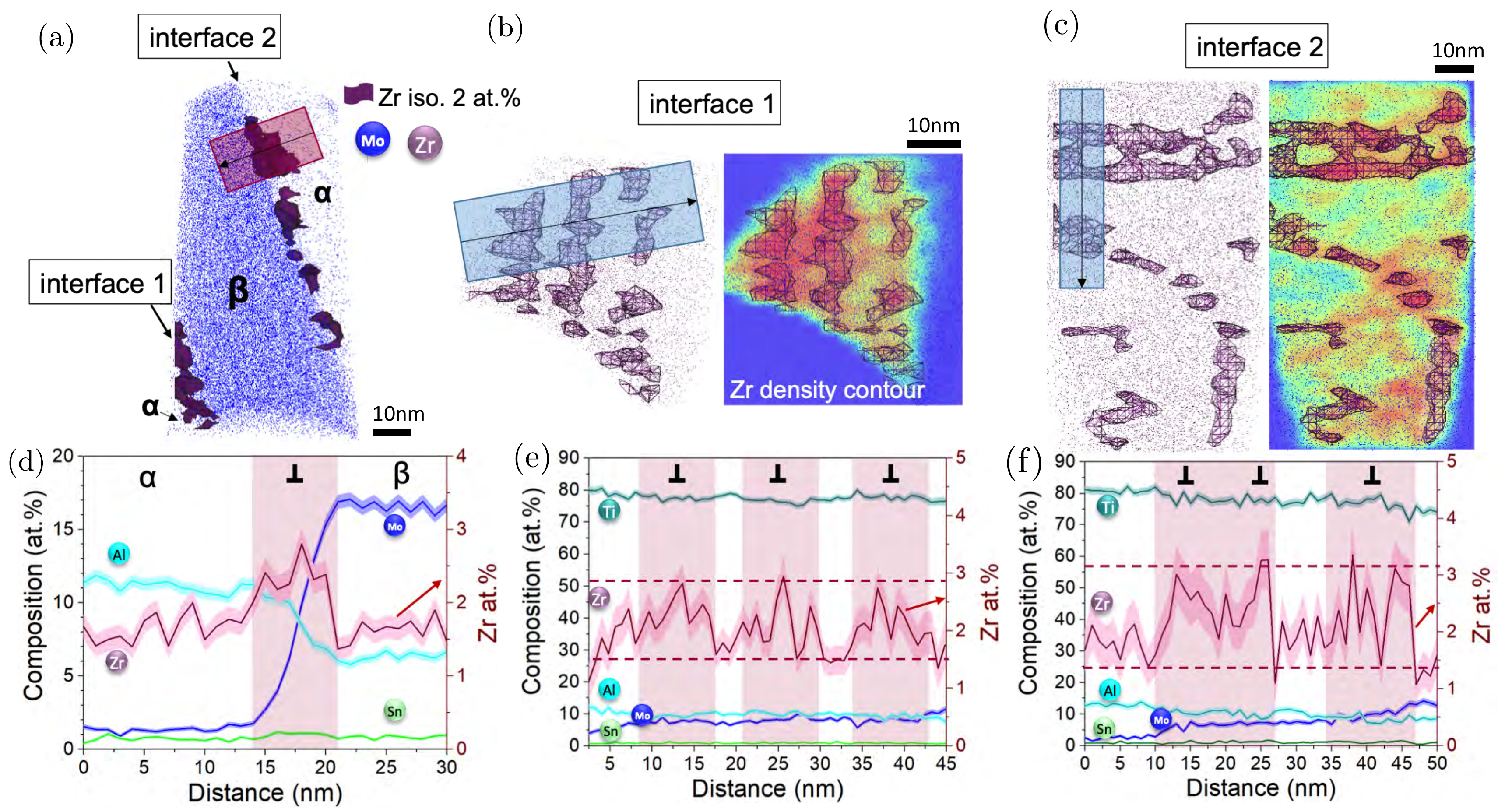}
\caption{Atom probe tomography of a specimen of fully heated treated Ti-6246. (a) Two $\alpha$ lamellae were observed within the needle, with the $\beta$ matrix in-between, as labelled. The point cloud represents a volume of $50\usk nm \times 50\usk nm \times 95\usk nm$. (b-c) 2 at.$\%$ Zr isosurface highlighting regions of increased Zr concentration at each $\alpha/\beta$ interface. A density contour highlights regularly spaced regions along the interface with higher concentrations of Zr, which are attributed to possible defects along the interface. (d-f) Concentration profile across the regions highlighted in (a-c), respectively.}
\label{fig:APT}
\end{figure*}

The chemistry of a similar interface was investigated using APT, Figure~\ref{fig:APT}. When a 2 at.$\%$ Zr isosurface is applied, Figure~\ref{fig:APT}(a), regularly spaced features can be seen along both of the $\alpha/\beta$ interfaces examined, labelled interface 1 and 2. To further highlight this, a density plot was used, figure~\ref{fig:APT}(b-c). These analyses within the plane of the interface make clear that there are regularly spaced enriched regions of Zr along the interface, which form linear features.

When composition profiles are taken (along the directions indicated by arrows in each sub-figure), an increase of Zr of around 1.5 at. $\%$ is measured, spaced around 5--10 nm apart. These are observed to occur in both interfaces examined, at multiple locations and in multiple datasets obtained from different specimens prepared from the same sample, and is hence considered to be a general feature of such interfaces. 

\subsection{Deformed $\alpha$/$\beta$ interface}
\begin{figure}[h!]
\centering\includegraphics[width=1\linewidth]{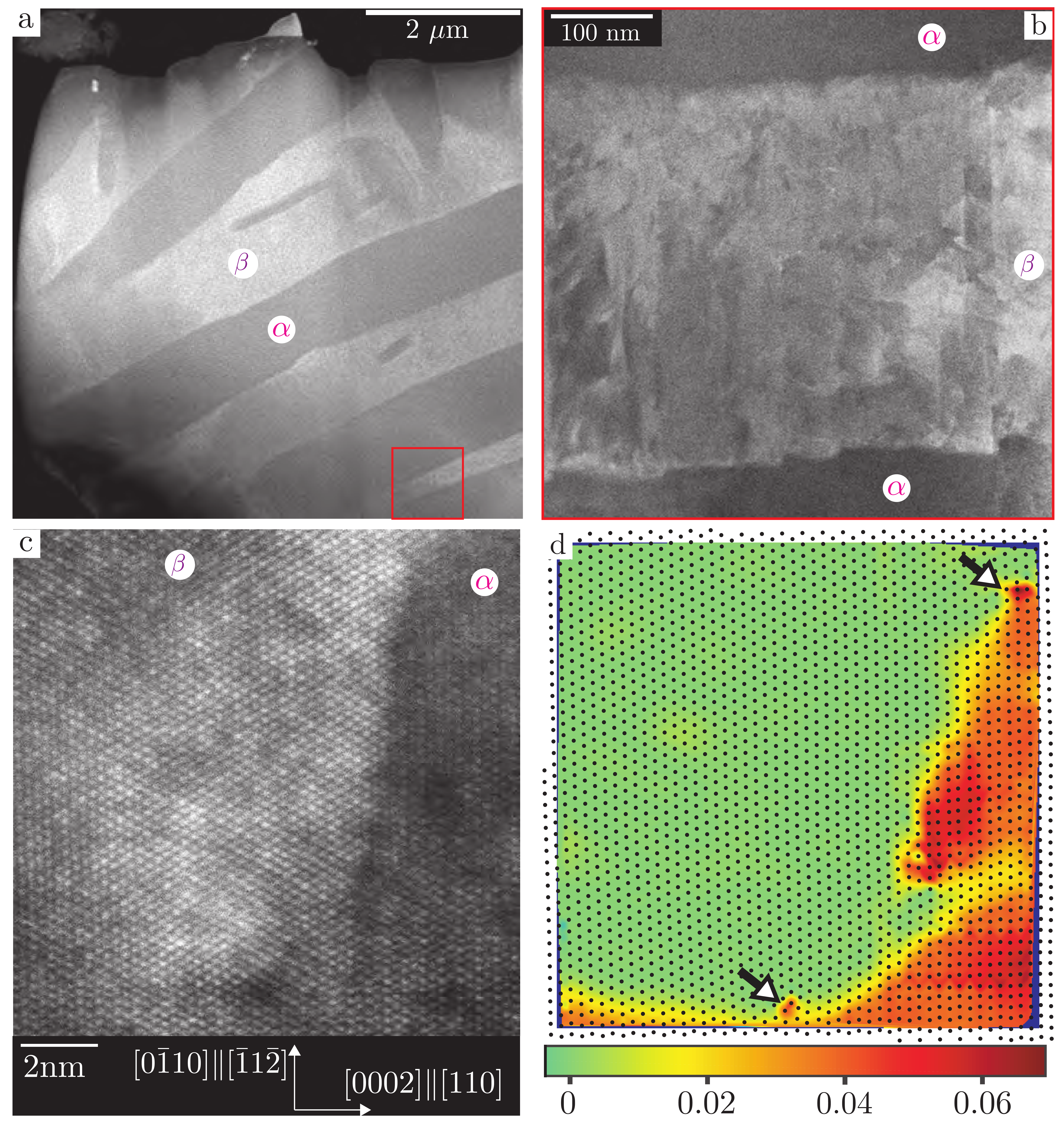}
\caption{(a) HAADF-STEM overview of the TEM foil  (image normal $[1\bar{2}\bar{1}0]_{\alpha}\|[1\bar{1}\bar{1}]_\beta$), with (b) a magnified image of the $\alpha$ plate structure. The $\alpha$ phase appears as dark and the $\beta$ phase as a light grey. Large steps can be seen along the $\alpha$/$\beta$ interface, and evidence of distortion can be seen in the $\beta$ matrix. (c) shows the increased magnification of the less coherent step structure and (d) is a colour map of the centre of symmetry parameter~\cite{Kelchner1998}, with dislocations highlighted by arrows.}
\label{fig:DFalpha}
\end{figure}

The Ti-6246 specimen deformed to 10$\%$ strain is shown in Figure~\ref{fig:DFalpha}, where the image plane is parallel to (111)$_{\beta}$//(2$\overline{11}$0)$_{\alpha}$. The $\alpha$ laths are less lenticular than in the undeformed alloy, with characteristic `bumps' along the interface. Slip bands can be seen in the $\beta$ matrix, terminating at large heterogeneous steps along the $\alpha$/$\beta$ interface. Within the $\beta$ phase, the  trace of the slip bands can be seen on $\{110\}$ glide planes, which lie parallel to the beam direction.  

Given the symmetry of the $\beta$ and $\alpha$ lattices, a hexagonal centre-of-symmetry parameter~\cite{Kelchner1998} was calculated for each atomic column, Figure~\ref{fig:DFalpha}(d). The parameter is zero in a perfect $\beta$ crystal, but takes on high values in the $\alpha$, which possess off-hexagonal symmetry when viewed along $[2\overline{1}\overline{1}0]_{\alpha}$ . A gross rotation in the crystal lattice can be seen in the image. Given the numerous dislocations observed in Figure ~\ref{fig:DFalpha}, their pile-up at the interface is the likely cause of the $\beta$ lattice rotation. The $\beta$ phase is softer than the $\alpha$ phase, and therefore might be expected to accommodate more of the applied deformation. Some lattice dislocations in the $\beta$ phase are evident in Figure~\ref{fig:DFalpha}(d) and are labelled with arrows.

Figure~\ref{fig:NGalpha} shows clearly that steps in the $\alpha$/$\beta$ interface allow for the accommodation of lattice mismatch between the two phases in Ti-6246. The interfacial steps become significantly exaggerated when the alloy is cold worked, Figure~\ref{fig:DFalpha}. One might hypothesize that these could act as nucleation sites for the formation of secondary $\alpha$. Conversely, the large shear bands, seen in Figure~\ref{fig:DFalpha}, might also provide nucleation sites at lower ageing temperatures. 

\section{Discussion}

Zherebtsov \emph{et al.}~\cite{Zherebtsov2010},  investigated the loss of coherency at stepped $\alpha$/$\beta$ interphase boundaries using the analytical van der Merwe model. The study showed that the absorption of lattice dislocations by the interface leads to gradual loss of coherency substantially increasing the interfacial energy. The minimisation of this energy is likely to drive the formation of the giant interfacial steps in deformed Ti-6246 observed in this study.   

Formation of interfacial phases has been suggested to accompany the $\beta$ to $\alpha$ solid state transformation in some Ti alloys, helping to accommodate the interfacial misfit strain and  potentially providing nucleation sites for the $\alpha$ plates. Kang \emph{et al.}~\cite{Kang2000} investigated the $\alpha$/$\beta$ phase boundary in a near-$\alpha$ titanium alloy, observing a discontinuous interface phase in a pre-strained material. Nag \emph{et al.}~\cite{Nag2009} reported assisted nucleation of $\alpha$ laths from aged-in nanoscale $\omega$ phase in Ti-5553-0.5Fe, which can be exploited for microstructural refinement~\cite{Zheng2016}. Our study has not found any additional phases associated with the interphase boundary. Instead interfacial strain relief is provided by a $1\usk$nm thick region where the $\alpha$ lattice shows progressive distortion toward the \emph{bcc} structure.   

Considering the thermodynamics of the interface, deformation has resulted in an accumulation of defects and associated interfacial defect structures which are available for energy minimisation by thermally-induced diffusional rearrangement. Many titanium alloys such as Ti-6246, Ti-6Al-4V and Ti-6Al-2Sn-4Zr-2Mo are worked in the two phase region, after which a heat treatment is applied to remodel the primary $\alpha$, termed globularisation. Jackson \emph{et al.}~\cite{Jackson2009} showed in the near-$\beta$ alloy Ti-10V-2Fe-3Al that deformation causes the $\alpha$ to fragment, a precursor to globularisation. Dislocation pile-up occurs at the $\alpha$/$\beta$ boundary causing the break up of $\alpha$ plates, leading to penetration of $\beta$ at the subgrain boundaries. Such dissolution of $\alpha$ by $\beta$ phase must therefore be driven by the diffusion of highly misfitting atoms such as Mo, which tend to be $\beta$-stabilisers, to highly defective regions such as those observed here. This provides a mechanistic hypothesis for how globularisation occurs. Additionally, Poschmann \emph{et al.}~\cite{Poschmann2018} found \emph{via} elasticity theory and molecular dynamics that $<111>_{\beta}$ screw dislocations can act as heterogeneous nucleation sites in pure Ti, resulting in an elastically preferential habit plane. This indicates that after deformation, $\alpha$ nucleation that occurs may be affected by the Burgers vector of the dislocation, linking dislocation density to the post transformation microstructure. Therefore the prevalence of $\{110\}_\beta$ slip bands observed here could lead to altered nucleation of secondary $\alpha$.  It should also be noted that slip bands through the $\alpha-\beta$ ensemble provide both a means of work hardening and of flow localisation, which might be of concern, \emph{e.g.} in fatigue. Such slip transfer events have been observed and analysed in detail elsewhere~\cite{Joseph2018}, and therefore will not be discussed further.

Elemental segregation at the $\alpha/\beta$ interface has not yet been seen in the literature. The Zr peaks observed could be due to defects along the interface\textcolor{red}{, however the spacing observed (5-10 nm) is inconsistent with the occurrence of these suggested defects at every interface step.} Selective segregation patterns associated with the local strain state were recently reported in a detailed analysis by a combination of HR-STEM and APT~\citep{Liebscher2018}. For the APT data presented here, the zone axis along which these compositional variations were observed is not known. Therefore it cannot directly be connected crystallographically to the features seen in the HR-STEM samples. Zr is a relatively neutral stabiliser in Ti alloys, segregating to neither the $\alpha$ nor the $\beta$ phase. However, the segregation of Zr was predicted to help lower the stacking fault energy in $\alpha$-Ti ~\cite{Kwasniak2016}, so segregation hence likely also helps minimise the system's free energy. The spacing of the Zr-rich observed features is similar to that of the slip bands observed after deformation in the $\beta$ matrix, \emph{e.g.} before the secondary $\alpha$ plates observed here were grown. Ultimately this can not be known without a fully-correlative, in-situ observation of the slip bands and then ageing of the $\alpha$, \emph{e.g.} by HR-STEM, followed by correlative APT to detect the Zr features. It is however reasonable to state, based on similar observations in the literature~\citep{Liebscher2018, Araullo-Peters2012,KwiatkowskidaSilva2018}, that the pattern in the Zr segregation at the interface is due to the presence of line defects.

Titanium plates are held to thicken by ledgewise growth, since the interface steps will be high energy features with consequent high mobility. This gives them the ability to `sweep up' solutes and other interfacial defects as they move. This may provide a rationale for why such segregation might occur. It may also be assisted by Zr transport along slip bands, provided that there is a thermodynamic driving force. However, the mobility of such interface defects would then be kinetically inhibited by the energy required to un-pin from the solute, and/or the requirement for a waiting time for the solute to diffuse to the moving defect. This would ultimately limit the ability of secondary $\alpha$ to coarsen, \emph{e.g.} during elevated temperature service, acting to stabilise the microstructure in use. It would also limit the ability of such deformation generated mobile dislocations to interact with existing interfacial conherency defects, thus strengthening the interface.

\section{Conclusions}
We examined the $\alpha$/$\beta$ interface in Ti-6Al-2Sn-4Zr-6Mo (Ti-6246). Semi-coherent interface steps are observed at a spacing of 4.5 $\pm$1.13 atoms and an inclination, $\omega$ of 12.6$\pm1\degree$, in the as-grown condition which is in good agreement with theory predictions, where the topological model predicted a disconnection spacing $\lambda_T$ of 4.25 atoms (1.22$\usk$nm), with $\omega_T$=12.29$^{\circ}$ and $\phi_T$=0.5337$^{\circ}$ and the PTMC model predicted an angle of $\omega_P$=$12.27^\circ$ and $\phi_P$=0.5331$^{\circ}$. A $\sim1\usk\nano\meter$ interface region is observed where the two lattices are distorted.  Deformed $\alpha$ exhibited larger, less coherent steps with slip bands lying in $\{110\}_\beta$. These microstructural steps may provide the sources for the pinch-off process of globularisation observed during the $\alpha-\beta$ heat treatment of $\alpha+\beta$ worked titanium alloys. Furthermore,  Zr decoration of the interface was observed, localised within linear features with a spacing of 5-10 nm.

\section*{Acknowledgements}
The authors acknowledge funding from EPSRC (grants EP/K034332/1 and EP/M506345/1 and EP/L025213/1), Rolls-Royce plc and Imperial College. The assistance of Prof. Robert C. Pond was of great help in improving the manuscript, particularly the mathematical treatment. Uwe Tezins and Andreas Sturm for their technical support of the atom probe tomography and focused ion beam facilities at the Max-Planck-Institut für Eisenforschung. L.T. Stephenson is acknowledged for his kind help and support for the cryo-FIB and overall setup of the Laplace project. Dr Yanhong Chang is grateful to the China Scholarship Council (CSC) for funding of her PhD scholarship. BG acknowledges financial support from the ERC-CoG-SHINE-771602 and is grateful for the financial support from the BMBF via the project UGSLIT and the Max-Planck Gesellschaft via the Laplace project. 

\bibliography{Interface-2}
\end{document}